\documentclass{article}

\usepackage{comment}

 \usepackage[preprint,nonatbib]{neurips_2019}

\usepackage[utf8]{inputenc} 
\usepackage[T1]{fontenc}    
\usepackage{hyperref}       
\usepackage{url}            
\usepackage{booktabs}       
\usepackage{amsfonts}       
\usepackage{nicefrac}       
\usepackage{microtype}      
\usepackage{todonotes}
\usepackage{etoolbox}
\usepackage{amsmath}

\newtoggle{anonymous}
\newcommand{\Zygote}{\iftoggle{anonymous}{$\partial$P.jl}{Zygote}}
\newcommand{\Zygoteplain}{\iftoggle{anonymous}{$\partial$P.jl}{Zygote}}

\title{ $\partial P$: A Differentiable Programming   System to Bridge Machine Learning and Scientific Computing}

\author{
  Mike Innes\thanks{mike@juliacomputing.com} \\   
  Julia Computing
  \And
  Alan Edelman \\
  MIT
  \And
  Keno Fischer \\
  Julia Computing
  \And
  Chris Rackauckas \\
  MIT, UMB
  \And
  Elliot Saba \\
  Julia Computing
  \And
  Viral B. Shah \\
  Julia Computing
  \And
  Will Tebbutt \\
  Invenia Labs
}

\begin{document}

\maketitle

\begin{abstract}
    Scientific computing is increasingly incorporating the advancements in machine learning and the ability to work with large amounts of data. At the same time, machine learning models are becoming increasingly sophisticated and exhibit many features often seen in scientific computing, stressing the capabilities of machine learning frameworks. Just as the disciplines of scientific computing and machine learning have shared common underlying infrastructure in the form of numerical linear algebra, we now have the opportunity to further share new computational infrastructure, and thus ideas, in the form of Differentiable Programming. 
    
    We describe a Differentiable Programming ($\partial P$) system that is able to take gradients of Julia programs
    making Automatic Differentiation  a first class language feature.
    Our system supports almost all language constructs (control flow, recursion, mutation, etc.) and compiles high-performance code without requiring any user intervention or refactoring to stage computations. This enables an expressive programming model for deep learning and, more importantly, it enables users to utilize the existing Julia ecosystem of scientific computing packages in deep learning models.
    
    We discuss support for advanced techniques such as mixed-mode, complex and checkpointed differentiation, and present how this leads to efficient code generation. We then showcase several examples of differentiating programs and mixing deep learning with existing Julia packages, including differentiable ray tracing, machine learning on simulated quantum hardware, training neural stochastic differential equation representations of financial models and more.
\end{abstract}

\section{Introduction}

    At first glance, a casual practitioner might think that scientific computing and machine learning are different scientific disciplines. Modern machine learning has made its mark through breakthroughs in neural networks. Their applicability towards solving a large class of difficult problems in computer science has led to the design of new hardware and software to process extreme amounts of labelled training data, while simultaneously deploying trained models in devices. Scientific computing, in contrast, a discipline that is perhaps as old as computing itself, tends to use a broader set of modelling techniques arising out of the underlying physical phenomena. Compared to the typical machine learning researcher, many computational scientists works with smaller volumes of data but with more computational complexity and range. As we look closer, many similarities emerge. Both disciplines have a preference for using dynamic languages such as Python, R and Julia. Often, performance critical sections in Python and R are written in C++ and Fortran, less so in Julia. The core computational routines are grounded in numerical linear algebra, and fundamentally, hardware has been designed to accelerate this computational core. 

    There are deep historical and intellectual links between machine learning and scientific computing.  On the surface, deep learning appears to be data driven and scientific computing is about very technical numerically intensive differential equations that mirror physical processes. It is our viewpoint that the fields are closer than is often realized with opportunities for machine learning and scientific computing to benefit from stronger interactions. 
    
    Specifically, machine learning has strongly benefited from {\bf Numerical Linear Algebra} software. The optimized numerical linear algebra stack was first developed in the context of scientific computing - BLAS (Levels 1 through 3) kernels~\cite{blas}, LAPACK routines for computing matrix factorizations~\cite{lapack}, and MPI for message passing in parallel computing~\cite{mpi}. The early CUDA libraries were developed by Nvidia for accelerating scientific kernels on GPUs.

    This trend is continuing with more recent techniques such as {\bf Neural ODEs}. A neural ODE \cite{chen2018neural} is a neural network layer which compacts the $L$ layers of a ResNet into a single ODE definition solved in $N<L$ steps of an adaptive ODE solver. By changing to this form, the memory and computational costs for the residual network model are decreased.  Additionally, the continuous ODE formulation does not restrict the model to a grid like the traditional RNN, naturally allowing for fitting time series with irregular observations. Existing ODE solver software can then be employed for efficient solution of the system.
    
    This is a two-way street. The 2018 Gordon Bell prize~\cite{gordon-bell-2018-a}, awarded for the largest high-performance scientific computing application, applied deep learning to climate analytics at exascale. Scientific computing also has much to benefit from advances in machine learning:
    
    \begin{enumerate}
    \item{\bf Surrogate modeling:}  Scientific simulations are often expensive to run as they evaluate a system using first principles. These simulations can be accelerated by having machine learning models approximate the input-output relation. Neural networks or other surrogate models can be trained on expensive simulations once and then used repeatedly in place of the simulations, making it possible to explore the parameter space, propagate uncertainties, and fit the data in ways that have previously been impossible.

    \item{\bf Adjoint sensitivity analysis:} Calculating the adjoint of an ordinary differential equation system $\frac{du}{dt} = f(u,p,t)$ requires solving the reverse ODE $\frac{d\lambda^{*}}{dt} = \lambda^{*} \frac{df}{du} + \frac{df}{dp}$. The term $\lambda^{*} \frac{df}{du}$ is the primitive of backpropogation, and thus applying machine learning AD tooling to the ODE function $f$ accelerates the scientific computing adjoint calculations.
    
    \item{\bf Inverse problems:}  For many parameterized scientific simulations, one can ask "what parameters would make my model best fit the data?" This inverse problem is pervasive yet difficult because it requires the ability to efficiently compute the gradient of a large existing simulation. One can train a model on a simulator, which can then be used to quickly solve inverse problems, but this currently requires generating massive amounts of simulation data for training, which is slow and computationally expensive. By being able to differentiate through simulators, we can learn much more quickly and efficiently.

    \item{\bf Probabilistic Programming:} Inference on statistical models is a crucial tool in the sciences. Probabilistic programming  
    enables more complex models and scaling to huge data sets by combining statistical methods with the generality of programming language constructs. Automatic differentiation is the backbone of many probabilistic programming tools, but domain specific languages lack access to an existing ecosystem of tools and packages. $\partial P$ in a general purpose language has the benefit of higher composability, access to better abstractions, and enabling richer and more accurate models. The Turing.jl~\cite{Turing.jl-2018} and Gen.jl~\cite{Gen.jl-2019} packages are excellent examples of these capabilities.

\end{enumerate}

    Differentiable Programming ($\partial P$) has the potential to be the lingua franca that can further unite the worlds of scientific computing and machine learning. The choice of a language to implement this system is an important one. Supporting multiple languages within a single $\partial P$ system causes an explosion in complexity, vastly increasing the developer effort required. Our $\partial P$ system extends the Julia programming language~\cite{Julia-2017-a} with differentiable programming capabilities. We chose the Julia language because of the abundance of pure-Julia packages for both machine learning and scientific computing allowing us to test our ideas on fairly large real-world applications.
    
    Our system can be directly used on existing Julia packages, handling  user-defined types, state-based control flow, and plentiful scalar operations through source-to-source AD. In this paper we briefly describe how we achieve our goals for a $\partial P$ system and showcase its ability to solve problems which mix machine learning and pre-existing scientific simulation packages. 
    
\subsection{A simple {\tt sin} example: Differentiate Programs not Formulas}

We start out with a very simple example to differentiate $\sin(x)$ written as a program through its Taylor series:
$$\sin x =
x - \frac{x^3}{3!}
+\frac{x^5}{5!}-\ldots .$$

One feature of our example is that the number of terms will not be fixed, but will depend on x through a numerical convergence criterion.

To run, install Julia v1.1 or higher, and install the Zygote.jl and ForwardDiff.jl packages with:

\begin{verbatim}
using Pkg
Pkg.add("Zygote")
Pkg.add("ForwardDiff")
using Zygote, ForwardDiff


function s(x) 
    t = 0.0
    sign = -1.0 
    for i in 1:19   
        if isodd(i)
            newterm = x^i/factorial(i)
            abs(newterm)<1e-8 && return t
            println("i=",i)
            sign = -sign
            t += sign * newterm
        end
    end
    return t 
end
\end{verbatim}

While the Taylor series for sine could have been written more compactly in Julia, for purposes of illustrating more complex programs, we purposefully used a loop, a conditional, a print statement, and function calls to \verb+isodd+ and \verb+factorial+, which are native Julia implementations. AD just works, and that is the powerful part of the Julia approach. Let’s compute the gradient at x = 1.0 and check whether it matches \verb+cos(1.0)+:

\begin{verbatim}
julia> ForwardDiff.derivative(s, 1.0) # Forward Mode AD
i=1
i=3
i=5
i=7
i=9
i=11
0.540302303791887

julia> Zygote.gradient(s, 1.0) # Reverse Mode AD
i=1
i=3
i=5
i=7
i=9
i=11
(0.5403023037918872,)

julia> cos(1.0)
0.5403023058681398 
\end{verbatim}

\section{Implementation}
\label{Impl}

    Recent progress in tooling for automatic differentiation (AD) has been driven primarily by the machine learning community. Many state of the art reverse-mode AD tools such as Tracker.jl~\cite{Flux.jl-2018,2019sbc}, PyTorch~\cite{pytorch}, JAX~\cite{jax}, and TensorFlow~\cite{tensorflow} (in the recent Eager version) employ tracing methods to extract simplified program representations that are more easily amenable to AD transforms. These traces evaluate derivatives only at specific points in the program space. Unfortunately, this generally unrolls all control flow and requires compilation and optimization for every new input value.
    
    This choice has been driven largely by the fact that, as the JAX authors put it, ``ML workloads often consist of large, accelerable, pure-and-statically-composed (PSC) operations''~\cite{jax}. Indeed, for many ML models the per-executed-operation overhead (in both time and memory) incurred by tracing-based AD systems is immaterial, because these execution time and memory requirements of the operations dwarf any AD overhead.
    
    However, this assumption does not hold for many scientific inverse problems, or even the cutting edge of ML research. Instead, these problems require a $\partial P$ system capable of: (1) low overhead, independent of the size of the executed operation (2) Efficient support for control flow (3) Complete, efficient support for user defined data types (4) Customizability (5) Composability with existing code unaware of $\partial P$, and (6) Dynamism.

    Particularly, scientific programs tend to have adaptive algorithms, whose control flow depends on error estimates and thus the current state of the simulation, numerous scalar operations, define large nonlinear models using every term individually or implementing specialized numerical linear algebra routines, and pervasive use of user-defined data structures to describe model components, which require efficient memory handling (stack-allocation) in order for the problem to be computationally feasible. 

    To take these kinds of problems, \Zygote\ does not utilize the standard methodology and instead generates a derivative function directly from the original source which is able to handle all input values. This is  called a source-to-source transformation, a methodology with a long history~\cite{baydin2018automatic} going back at least to the ADIFOR source-to-source AD program for FORTRAN 77~\cite{adifor}. Using this source-to-source formulation, \Zygote\ can then be compile, heavily optimize, and re-use a single gradient definition for all input values. Significantly, this transformation keeps control flow in tact: not unrolling loops to allow for all possible branches in a memory-efficient form. However, where prior source-to-source AD work has often focused on static languages, \Zygote\ expands upon this idea by supporting a full high level language, dynamic, Julia, in a way that allows for its existing scientific and machine learning package ecosystem to benefit from this tool. 

\subsection{Generality, Flexibility, and Composability}

    One of the primary design decisions of a $\partial P$ system is how these capabilities should be exposed to the user. One convenient way to do so is using a differential operator $\mathcal{J}$ that operates on first class functions and once again returns a first class function (by returning a function we automatically obtain higher order derivatives, through repeated application of $\mathcal{J}$). There are several valid choices for this differential operator, but a convenient choice is

\[
    \mathcal{J}(f) := x \to (f(x), J_f(x)z),
\]

i.e. $\mathcal{J}(f)(x)$ returns the value of $f$ at x, as well as a function which evaluates the jacobian-vector product between $J_f(x)$ and some vector of sensitivities $z$. From this primitive we can define the gradient of a scalar function $g: \mathbb{R}^n \to \mathbb{R}$ which is written as:

\[
    \nabla g(x) := \left[\mathcal{J}(g)(x)\right]_2(1)
\]

    ($[]_2$ selects the second value of the tuple, $1 = \partial z / \partial z$ is the initial sensitivity).

    This choice of differential operator is convenient for several reasons: (1) The computation of the forward pass often computes values that can be re-used for the computation of the backwards pass. By combining the two operations, it is easy to re-use this work. (2) It can be used to recursively implement the chain rule (see figure \ref{chain}).

\begin{figure}
    \begin{minipage}{0.5\textwidth}
    \begin{verbatim}
    function J(f . g)(x)
        a, da = J(f)(x)
        b, db = J(g)(a)
        b, z -> da(db(z))
    end
    \end{verbatim}
    \caption{The differential operator $\mathcal{J}$ is able to implement the chain rule through a local, syntactic recursive transformation.}\label{chain}
    \end{minipage}
    \hspace{0.1in}
    \begin{minipage}{0.5\textwidth}
    \begin{verbatim}
    julia> f(x) = x^2 + 3x + 1
    julia> gradient(f, 1/3)
    (3.6666666666666665,)
    
    julia> using Measurements;
    julia> gradient(f, 1/3 +- 0.01)
    (3.6666666666666665 +- 0.02,)
    \end{verbatim}
    \caption{With two minimal definitions, \Zygote is able to obtain derivatives of any functions that only requires those definitions, even through custom data types (such as \textit{Measurement}) and many layers of abstraction.}\label{generalize}
    \end{minipage}
\end{figure}

    This second property also suggests the implementation strategy: hard code the operation of $\mathcal{J}$ on a set of primitive $f$'s and let the AD system generate the rest by repeated application of the chain rule transform. This same general approach has been implemented in many systems \cite{pearlmutter2008reverse,wang2018demystifying} and a  detailed description of how to perform this on Julia's SSA form IR is available in earlier work \cite{Zygote.jl-2018}.

    However, to achieve our extensibility and composability goals, we implement a slight twist on this scheme. We define a fully user extensible function $\partial$ that provides a default fallback as follows
\[
\partial(f)(args...) = \mathcal{J}(f)(args...),
\]
    where the implementation that is generated automatically by $\mathcal{J}$ recurses to $\partial$ rather than $\mathcal{J}$ and can thus easily be intercepted using Julia's standard multiple dispatch system at any level of the stack. For example, we might make the following definitions:

\[
\partial(f)(::\text{typeof}(+))(a::\text{IntOrFloat}, b::\text{IntOrFloat}) = a + b, z \to (z, z) \]\[
\partial(f)(::\text{typeof}(*))(a::\text{IntOrFloat}, b::\text{IntOrFloat}) = a * b, z \to (z * b, a * z)
\]

    i.e. declaring how to compute the partial derivative of $+$ and $*$ for two integer or float-valued numbers, but simultaneously leaving unconstrained the same for other functions or other types of  values (which will thus fall back to applying the AD transform). With these two definitions, any program that is ultimately just a composition of `+`, and `*` operations of real numbers will work. We show a simple example in figure \ref{generalize}. Here, we used the user-defined \textit{Measurement} type from the \textit{Measurements.jl} package~\cite{Measurements.jl-2016}. We did not have to define how to differentiate the $\wedge$ function or how to differentiate $+$ and $*$ on a \textit{Measurement}, nor did the \textit{Measurements.jl} package have to be aware of the AD system in order to be differentiated. This extra, user-extensible layer of indirection has a number of important consequences:

    \begin{itemize}
        \item \textbf{The AD system does not depend on, nor require any knowledge of primitives on new types.} By default we provide implementations of the differentiable operator for many common scalar mathematical and linear algebra operations, written with a scalar LLVM backend and BLAS-like linear algebra operations. This means that even when Julia builds an array type to target TPUs \cite{XLA.jl-2018}, its XLA IR primitives are able to be used and differentiated without fundamental modifications to our system.
    \item \textbf{Custom gradients become trivial.} Since all operations indirect through $\partial$, there is no difference between user-defined custom gradients and those provided by the system. They are written using the same mechanism, are co-optimized by the compiler and can be finely targeted using Julia's multiple dispatch mechanism.
\end{itemize}

Since Julia solves the two language problem, its Base, standard library, and package ecosystem are almost entirely pure Julia. Thus, since our $\partial P$ system does not require primitives to handle new types, this means that almost all functions and types defined throughout the language are automatically supported by \Zygote, and users can easily accelerate specific functions as they deem necessary.

\section{$\partial P$ in Practice}

A more extensive code listing for these examples is available at the following URL: \iftoggle{anonymous}{Hidden for review - an anonymized version is included as supplemental material}{https://github.com/MikeInnes/zygote-paper}.

\subsection{Deep Learning}
\label{dl}

\Zygote\ is a flexible backend for calculating gradients of deep learning models. A typical example is shown here, where a recurrent architecture using LSTMs \cite{Hochreiter:1997:lstm} is used to learn Shakespeare.
The code sample below demonstrates many powerful elements of \Zygote, making use of several convenient Julia features in the process.  First, the defined model has no special data types within it to enable AD; the models are defined for forward-pass calculation only, with backwards-pass definitions only for basic building blocks such as BLAS operations and basic array manipulation. \Zygote\ is used to wrap the loss computation, explicitly denoting the bounds of the computation that should be differentiated to calculate the gradients imposed upon the model, but all other pieces of code (including the LSTM layer definition itself) are written without automatic differentiation in mind. This model executes on the CPU, GPU~\cite{Flux.jl-2018} and Google TPU architecture~\cite{XLA.jl-2018}, with little to no change.

\begin{verbatim}
    alphabet, Xs, Ys = load_data("shakespeare_input.txt")

    model = Chain(
        LSTM(length(alphabet), 128),
        LSTM(128, 128),
        Dense(128, length(alphabet)),
        softmax,
    )
    
    opt = ADAM(0.01)
    
    # Run through our entire dataset a few times
    for epoch_idx in 1:10,
        (x_batch, y_batch) in zip(Xs, Ys)
        
        # Calculate gradients upon the model for this batch of data,
        # summing crossentropy loss across each time index in this batch
        grads = |\Zygoteplain|.gradient(model) do model
            return sum(crossentropy.(model.(x_batch), y_batch))
        end

        # Update the model, then reset its internal LSTM states
        model = update!(opt, model, grads)
        Flux.reset!(model)
    end
\end{verbatim}

\Zygote\ provides an extremely low-overhead AD interface. By performing source-to-source transformation, there is little runtime overhead from the AD transformation, beyond the actual cost of computing the backwards pass \footnote{In practice \Zygote\ may currently generate code that pessimizes certain compiler assumption and thus leads to slower execution. We are working on improving the compiler for these case to achieve true zero overhead.}. In addition, it has been shown to perform at the same level as TensorFlow for ResNet on a TPU pod~\cite{xla.jl}.

In order to measure this, we have benchmarked the backwards pass of a stacked LSTM network, measuring the runtime as batch size trends toward zero. We hypothesize that overall runtime should be linear in batch size, and so by linearly extrapolating our readings to a batch size of zero, we will be able to estimate the fixed overhead induced by each operation within the AD system, where an operation is a single adjoint definition.  We furthermore verify that this overhead is linear in the number of ops by measuring this for a variable number of stacked LSTMs and recording the number of ops per model.

Our benchmarks were run on an Intel(R) Core(TM) i5-6600 CPU @ 3.30GHz processor, running Linux with Julia version 1.3.  In all tests, multithreading was disabled both in the AD framework and in the underlying BLAS library, so as to avoid performance cliffs around multithreading thresholds. Our results are shown in Table \ref{table:zygote_overhead}, displaying an average overhead of \textbf{568.8 $ns$} per adjoint definition. This is a highly competitive as compared to other frameworks such as PyTorch, where $op$ overhead is at least 1 $\mu s$ \cite{PyTorchYear}.

\begin{table}[]
\centering
\begin{tabular}{cccc}
Layers & Total Overhead & Number of Ops & Overhead/op \\ \hline
1      & 147.0 us & 255 & 576.3 ns \\
2      & 280.5 us & 491 & 571.3 ns \\ 
3      & 406.1 us & 727 & 558.6 ns \\ \hline
\end{tabular}
\label{table:zygote_overhead}
\caption{\Zygote\ per-op overhead estimated across varying numbers of stacked LSTM layers.}
\end{table}

This vanishing overhead lowers the bar for AD systems to be integrated more thoroughly into programming languages at an extremely fine scale without worrying about performance issues. The lower the AD overhead, the smaller the minimum feasible kernel size for an AD system that is restricted by backwards pass efficiency.

\subsection{Differentiating a Trebuchet}

\begin{figure}[!htb]
  \centering\includegraphics[width=5.5in]{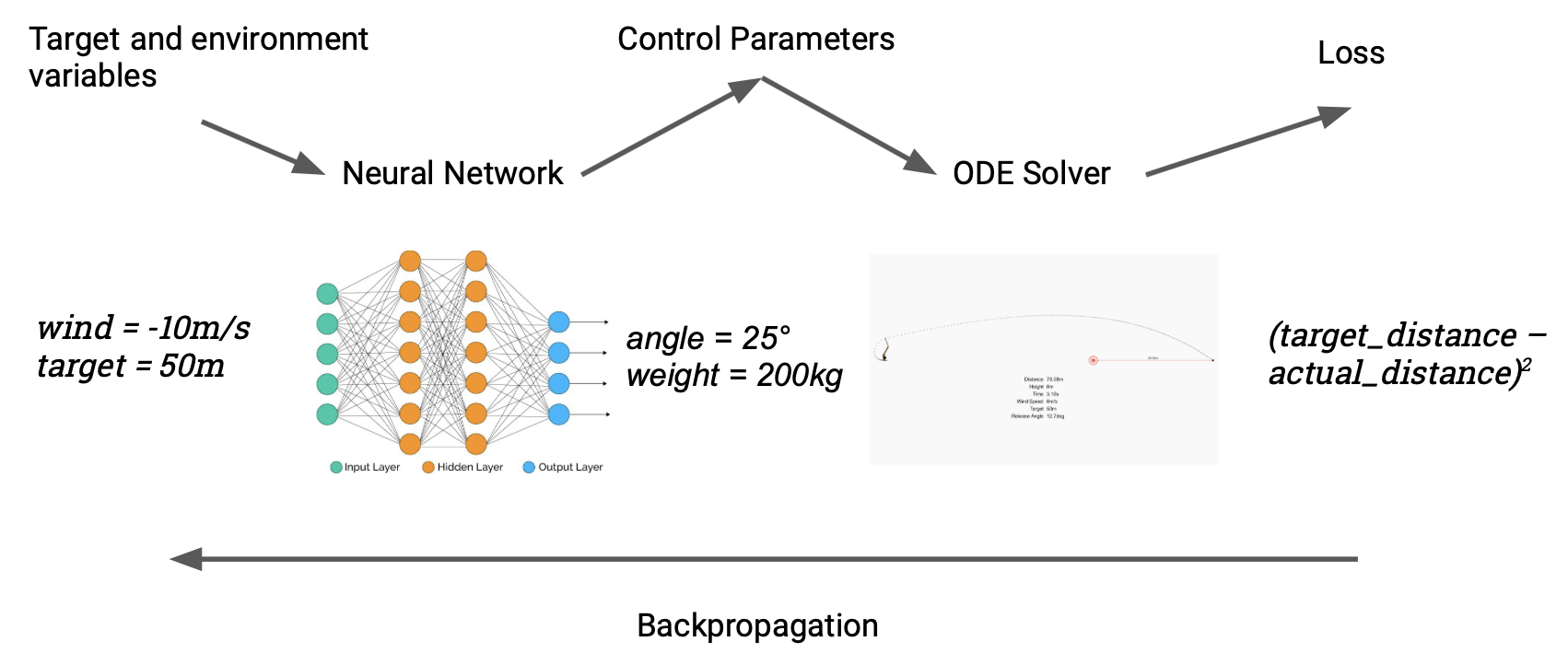}
  \caption{{\bf Using a neural network surrogate to solve inverse problems}}
  \label{fig:trebuchet}
\end{figure}

Model-based reinforcement learning has advantages over model-agnostic methods, given that an effective agent must approximate the dynamics of its environment \cite{atkeson1997comparison}. However, model-based approaches have been hindered by the inability to incorporate realistic environmental models into deep learning models. Previous work has had success re-implementing physics engines using machine learning frameworks \cite{Degrave_2019,de2018end}, but this effort has a large engineering cost, has limitations compared to existing engines, and has limited applicability to other domains such as biology or meteorology.

 \Zygote\ can be used for  control problems, incorporating the model into backpropagation with one call to \texttt{gradient}. We pick trebuchet dynamics as a motivating example. Instead of aiming at a single target, we optimize a neural network that can aim it given any target. The neural net takes two inputs, the target distance in metres and the current wind speed. The network outputs trebuchet settings (the mass of the counterweight and the angle of release) that get fed into a simulator which solves an ODE and calculates the achieved distance. We then compare to our target, and backpropagate through the entire chain to adjust the weights of the network. Our dataset is a randomly chosen set of targets and wind speeds. An agent that aims a trebuchet to a given target can thus be trained in a few minutes on a laptop CPU, resulting in a network which solves the inverse problem with constant-time aiming that is $100\times$ faster than optimizing the trebuchet system directly (Figure~\ref{fig:trebuchet}). We present the code for this and other common reinforcement learning examples such as the cartpole and inverted pendulum~\cite{rlvsdp}.

\subsection{Computer Vision}

An emerging direction for computer vision is to see the problem as `inverse graphics': where vision models take pixel arrays to scene parameters, renderers take scene parameters to pixel arrays. Many high-fidelity, photo-realistic renderers are available which contain a vast amount of implicit knowledge about this mapping, and differentiation allows renderers and models to be used in an autoencoder-like setup to quickly bootstrap a full vision model.

As in other cases, the difficulty lies in getting efficient derivatives from a production-ready renderer, typically written in a performance language like C++. We repeat the going themes of this paper: ML framework-based reimplementations are typically limited compared to existing renderers (both by the available framework operations and by the years of work from rendering experts that has gone into production renderers), and workarounds such as Monte Carlo sampling \cite{li2018differentiable} impose a high efficiency cost (around $40\times$ slower than a single render) compared to AD (at most around ~$5\times$ due to division, in principle). The Julia community's approach is to build a renderer suitable, first and foremost, for traditional production rendering tasks, using a general and high-performance numerical language, and then differentiate it. Of course, this approach does not preclude the use of domain-specific methods such as Monte Carlo sampling where appropriate (sections \ref{diffeq}, \ref{mixed}).

In our examples we have used our prototype renderer to demonstrate optimization of the position of a point light source, given the desired final rendered image. We define a loss function that accepts a point light source as input, renders the scene, and compares it to a reference image. As usual, gradients are then trivial and we can begin to update the position of the point source.

\begin{verbatim}
julia> guess = PointLight(Vec3(1.0), 20000.0, Vec3(1.0, 2.0, -7.0))

julia> function loss_function(light)
           rendered_color = raytrace(origin, direction, scene, light, eye_pos)
           rendered_img = process_image(rendered_color, screen_size.w,
                                        screen_size.h)
           return mean((rendered_img .- reference_img).^2)
       end

julia> gs = gradient(x -> loss_function(x, image), guess)   
\end{verbatim}

\begin{figure}[!htb]
\minipage{0.32\textwidth}
  \includegraphics[width=\linewidth]{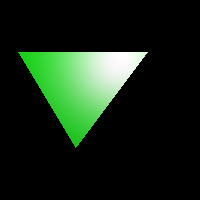}
  \caption{Initial Guess}\label{fig:raytracer1}
\endminipage\hfill
\minipage{0.32\textwidth}
  \includegraphics[width=\linewidth]{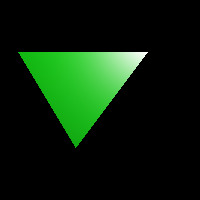}
  \caption{After 100 iterations}\label{fig:raytracer2}
\endminipage\hfill
\minipage{0.32\textwidth}%
  \includegraphics[width=\linewidth]{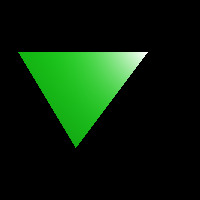}
  \caption{Target Image}\label{fig:raytracer3}
\endminipage
\end{figure}

\subsection{Financial Derivatives}

Much of finance is concerned with how the value of a security will change according to changes in the market factors, such as interest rates, market indices, or spot prices of underlying assets. This is naturally expressed via the derivatives of security prices with respect to various conditions. Previous work has noted that financial contracts are compositional, and form a kind of programming language~\cite{jones2000composing,jones2003write}. Work in Julia has generalized this to a differentiable contract language, Miletus~\cite{Miletus.jl-2019}, where greeks can b derived for arbitrary contracts. Contract valuation can be mixed with free-form market simulations as well as other $\partial P$ tools like neural networks, enabling differentiable programming for finance.

In working with fixed-income securities such as bonds, the primary factors are interest rate curves. These are typically quoted as the \emph{par rate} at a set of times in the future (e.g. 1 month, 3 month, 6 month, 1 year, etc.), which are referred to as "key rates". Using this to determine the price of a bond requires (1) interpolating a yield curve from the par rates: this is a process known as \emph{bootstrapping}, and requires solving a sequence of implicit problems, typically via Newton iterations; and (2) computing the bond price from the yield curve by discounting cash flows.

The derivatives of a bond price with respect to the key rates are known as \emph{key rate durations}: computing these via AD requires higher-order derivatives (in order to differentiate through the Newton solver). However, this is entirely invisible to the programmer; several unrelated uses of differentiation, and even entirely different ADs, simply compose together and do the right thing.

\subsection{Quantum Machine Learning}
\label{yao}

A promising approach for near-term exploitation of
quantum computing
is  hybrid classical-quantum algorithms where most  computation is performed on a classical computer, but key pieces are offloaded to a quantum processor. One approach to such hybrid systems is the \textit{variational quantum circuit} where a quantum circuit is parameterized by classical inputs. 
Here, a quantum circuit is parameterized by classical inputs and has a classical output corresponding to the expectation value of some observable of the final state (obtainable on real devices from the quantum measurement of the final state over repeated executions of the quantum computation).
One interesting characteristic of such quantum circuits is that it is generally possible to compute a (linear combination of) quantum circuit(s), the expectation value of which corresponds to the derivative of the expectation value of the original quantum circuit. An application is quantum circuit design, where the design phase the parameter space of the variational quantum circuit is explored to find a configuration that outputs desired values.

One such state of the art simulator is the Yao.jl~\cite{yao.jl} quantum simulator project. Yao.jl is implemented in Julia and thus composable with our AD technology. There are a number of interesting applications of this combination. The most obvious is to simply define a custom adjoint of the quantum evaluation that performs the quantum AD transform and evaluates the transformed circuit, thus enabling full end-to-end automatic differentiation of hybrid classical-quantum systems. This technique is of course applicable to both quantum simulators and real hardware.

A more subtle application is to simply perform traditional AD of the quantum simulator itself. As a simple example of this capability, we consider a Variational Quantum Eigensolver (VQE). A variational quantum eigensolver is used to compute the eigenvalue of some matrix $H$ (generally the Hamiltonian of some quantum system for which the eigenvalue problem is hard to solve classically, but that is easily encoded into quantum hardware). This is done by using a variational quantum circuit $\Phi(\theta)$ to prepare some quantum state $|\Psi\rangle = \Phi(\theta)|0\rangle$, measuring the expectation value $\langle \Psi|H|\Psi 0 \rangle$ and then using a classical optimizer to adjust $\theta$ to minimize the measured value. In our example, we will use a 4-site toy Hamiltonian corresponding to an anti-ferromagnetic Heisenberg chain:

\[
H = \frac{1}{4} \left[ \sum_{\langle i, j \rangle} \sigma_i^x \sigma_j^x + \sigma_i^y \sigma_j^y +  \sigma_i^z \sigma_j^z \right]
\]

\begin{figure}
    \begin{minipage}{0.48\textwidth}
    \begin{verbatim}
    using Yao, Zygote
    const nsites = 4
    let H = heisenberg(nsites),
        v0 = statevec(zero_state(nsites))
        energy(circuit) =
           (|$\Psi$| = circuit*v0;
            real(|$\Psi'$| * H * |$\Psi$|))
    end
    circuit_init =
        random_diff_circuit(nsites, 2)
    optimize_plot_loss(
        energy, circuit_init, ADAM(0.1))
    \end{verbatim}
  \caption{An ADAM optimizer is used to tune parameters of a variational quantum circuit to find the ground state of a 4-site anti-ferromagnetic Heisenberg chain Hamiltonian. The necessary gradients are obtained by automatic differentiation of a Yao.jl quantum simulator.}
\end{minipage}
\hspace*{.2in}
\minipage{0.48\textwidth}
  \includegraphics[width=\linewidth]{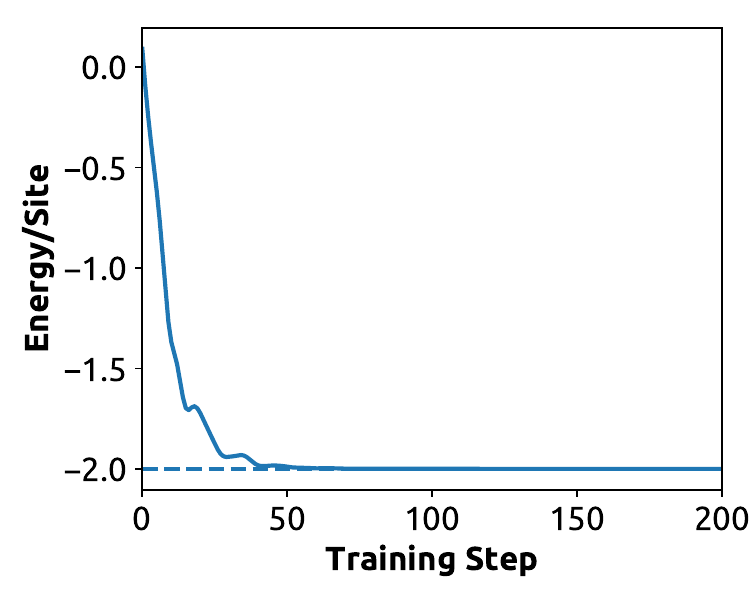}
  \caption{Optimization progress over steps of the classical optimizer to ground state.\label{fig:yao_loss}}
\endminipage
\end{figure}

We use a standard differentiable variational quantum circuit composed of layers (2 in our example) of (parameterized) rotators and $c_0$ entanglers with randomly initialized rotator angles. The corresponding code is showing in figure \ref{fig:yao_loss}  \footnote{For space efficiency, some details are factored into utility function omitted from this code listing. However, the code is very similar to the optimization loop from the deep learning example above. A full, expanded version of this example can be found with the full code listing in the supplemental material.}. The resulting plot can be seen in figure \ref{fig:yao_loss}.

\subsection{Neural Differential Equations with applications in finance}
\label{diffeq}

    Neural latent differential equations \cite{chen2018neural,pmlr-v5-alvarez09a,latent,DBLP:journals/corr/abs-1902-02376} incorporate a neural network into the ODE derivative function. Recent results have shown that many deep learning architectures can be compacted and generalized through neural ODE descriptions \cite{chen2018neural,He2016DeepRL,2019arXiv190401681D,DBLP:journals/corr/abs-1810-01367}. Latent differential equations have also seen use in time series extrapolation \cite{10.1093/bioinformatics/btn278} and model reduction \cite{ROMEROUGALDE2013170,8062736,2018arXiv180804930B,doi:10.1243/09544062JMES683}.

    Here we demonstrate financial time series forecasting with a neural stochastic differential equation (SDE). Typically, financial SDE models follow the form:

\[
    dX_t = f(X_t)dt + g(X_t)dW_t,
\]
    where
    $f:R^n \rightarrow R^n$ and $g:R^{n\times m} \rightarrow R^n$ with 
    $W_t$ as the $m$-dimensional Wiener process. For example, the infamous Black-Scholes equation

\[
    \frac{\partial V}{\partial t} + \frac{1}{2}\sigma^2 S^2 \frac{\partial^2 V}{\partial S^2} + rS\frac{\partial V}{\partial S} - rV = 0
\]

    is related through the Feynman-Kac Theorem to a probability measure driven by a geometric Brownian motion stock price $dS_t = r S_t dt + \sigma S_t dW_t$, where $S$ is the stock price, $r$ is the fixed interest rate of an option, and $\sigma$ is the volatility. This signifies that the true value of an option contract can then be achieved by hedging such that the following conditional expectation holds:

\[
    V(S,t) = \mathbb{E} \left[\int_t^T e^{-\int_t^T rd\tau}f(S_\nu)d\nu + e^{-\int_t^T rd\tau}\psi(S_T)d\nu    \right]
\]
    where $\psi(S_\nu)$ is the value of the contract at the terminal time $T$ given a stock price $S$, showing that the PDE's solution is given by an average over the SDE solutions. 

\begin{figure}[!htb]
  \centering\includegraphics[width=4in]{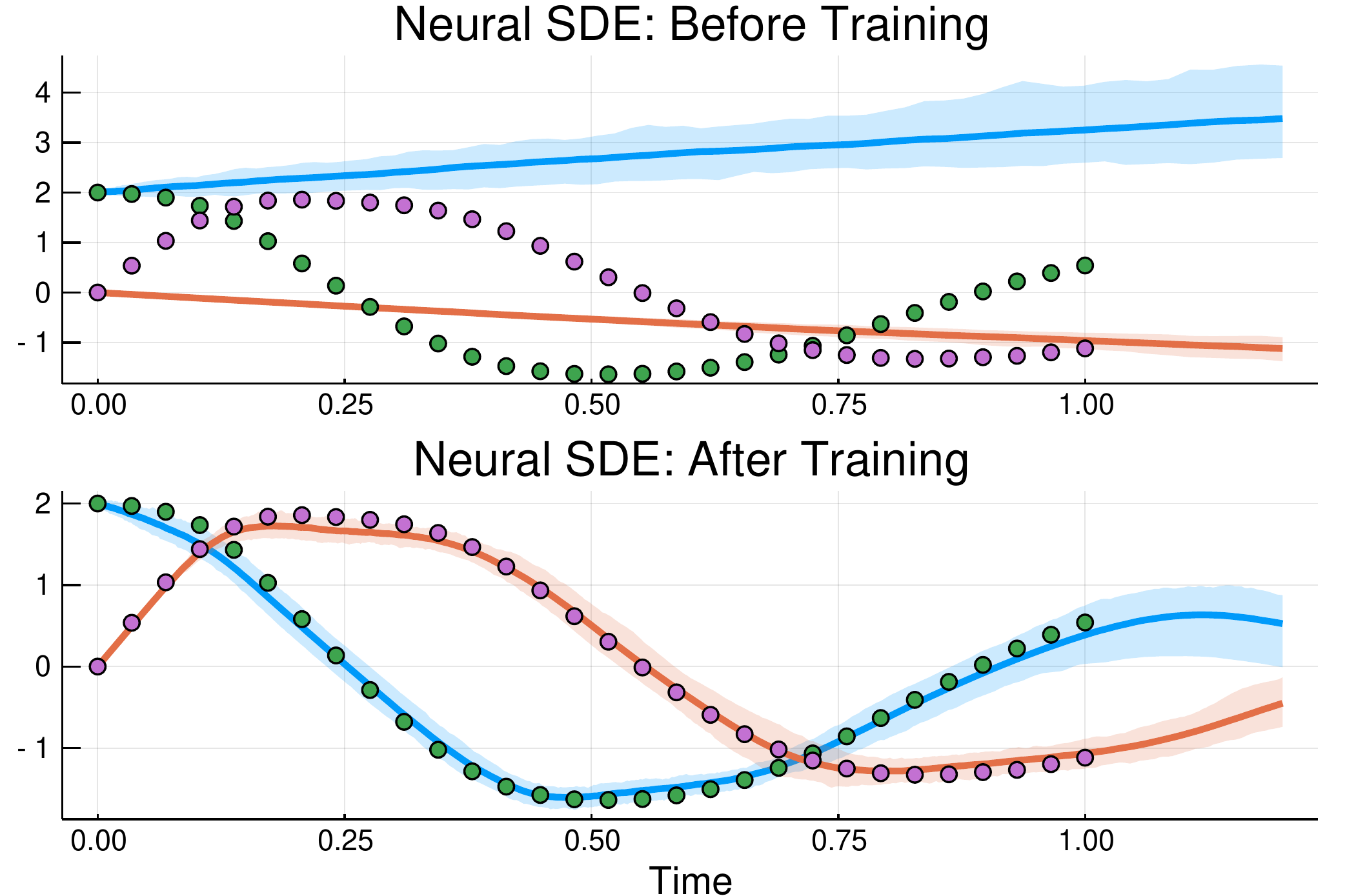}
  \caption{{\bf Neural SDE Training}. For the SDE solution $X(t)$, the blue line shows $X_1(t)$ while the orange line shows $X_2(t)$. The green points shows the fitting data for $X_1$ while the purple points show the fitting data for $X_2$. The ribbons show the 95 percentile bounds of the stochastic solutions.}\label{fig:neuralsde}
\end{figure}

    To generalize the model, we replacing the fixed interest rate $r$ with a neural network train against financial time series data. Our financial simulator utilizes a high strong order adaptive integration provided by DifferentialEquations.jl \cite{DifferentialEquations.jl-2017,Rackauckas2017ADAPTIVEMF}.  Figure \ref{fig:neuralsde} depicts a two-dimensional neural SDE trained using the $l_2$ normed distance between the solution and the data points. Included is a forecast of the neural SDE solution beyond the data to a final time of 1.2, showcasing a potential use case.

The analytical formula for the adjoint of the strong solution of a SDE is difficult to efficiently calculate due to the lack of classical differentiability of the solution\footnote{The the strong solution of stochastic differential equations are non-differentiable except on a measure zero set with probability 1. This means that standard Newtonian calculus cannot hold, and thus the derivation must take place in a the setting of a stochastic functional calculus known as the Malliavin Calculus \cite{doi:10.1137/S0363012902419059,huang2000malliavin,zhang_2004}.}. However, \Zygote\ still manages to calculate a useful derivative for optimization with respect to single solutions by treating the Brownian process as fixed and applying forward-mode automatic differentiation, showcasing \Zygote's ability to efficiently optimize its AD through mixed-mode approaches \cite{2018arXiv181201892R}. Common numerical techniques require computing the gradient with respect to a difference over thousands of trajectories to receive an average cost, while our numerical experiments suggest that it is sufficient with \Zygote\ to perform gradient decent on a neural SDE using single trajectories, reducing the overall computational cost by this thousands. This methodological advance combined with GPU-accelerated high order adaptive SDE integrators in DifferentialEquations.jl makes a whole new field of study accessible.

\section{Conclusion}

The disciplines of machine learning and scientific computing have much to share. We presented a $\partial P$ system that can serve as the basis of a common shared infrastructure in both disciplines. We demonstrated how we  can compose ideas in machine learning and scientific computing to allow for new applications that transcend these domains, by using the same technology to differentiate programs in both domains. On the ML side, we show the same performance as existing ML frameworks for deep learning models (on CPUs, GPUs, and TPUs) and in reinforcement learning. In the case of scientific computing, we show neural SDEs and quantum machine learning. The system is open source, and we invite the reader to try their own examples.

\subsubsection*{Acknowledgments}

This work would not have been possible without the work of many people. We are particularly indebted to our users that contributed the examples in this paper.Thanks to Kai Xu for the Bayesian colours example, Avik Pal for building our differentiable ray tracer, Tejan Karmali and Neethu Mariya Joy for the model-based RL experiments, Roger Luo and JinGuo Liu for quantum experiments, and Simon Byrne for the finance example. Thanks also to Zenna Tavares, Jesse Bettencourt and Lyndon White for being early adopters of \Zygote\ and its underlying technology, and shaping its development. Much credit is also due to the core Julia language and compiler team for supporting \Zygote's development, including Jameson Nash, Jeff Bezanson and others. Thanks also to James Bradbury for helpful discussions on this paper and many other things.

\small

\bibliography{zygote-dp}

\begin{thebibliography}{10}

\bibitem{mpi}
{\em MPI - A Message Passing Interface Standard}.
\newblock MPI Forum, 2015.

\bibitem{tensorflow}
M.~Abadi, P.~Barham, J.~Chen, Z.~Chen, A.~Davis, J.~Dean, M.~Devin,
  S.~Ghemawat, G.~Irving, M.~Isard, et~al.
\newblock Tensorflow: A system for large-scale machine learning.
\newblock In {\em 12th $\{$USENIX$\}$ Symposium on Operating Systems Design and
  Implementation ($\{$OSDI$\}$ 16)}, pages 265--283, 2016.

\bibitem{lapack}
E.~Anderson, Z.~Bai, C.~Bischof, S.~Blackford, J.~Dongarra, J.~Du~Croz,
  A.~Greenbaum, S.~Hammarling, A.~McKenney, and D.~Sorensen.
\newblock {\em LAPACK Users' guide}, volume~9.
\newblock {SIAM}, 1999.

\bibitem{atkeson1997comparison}
C.~G. Atkeson and J.~C. Santamaria.
\newblock A comparison of direct and model-based reinforcement learning.
\newblock In {\em Proceedings of International Conference on Robotics and
  Automation}, volume~4, pages 3557--3564. IEEE, 1997.

\bibitem{2018arXiv180804930B}
Y.~{Bar-Sinai}, S.~{Hoyer}, J.~{Hickey}, and M.~P. {Brenner}.
\newblock {Data-driven discretization: machine learning for coarse graining of
  partial differential equations}.
\newblock {\em arXiv e-prints}, page arXiv:1808.04930, Aug 2018.

\bibitem{baydin2018automatic}
A.~G. Baydin, B.~A. Pearlmutter, A.~A. Radul, and J.~M. Siskind.
\newblock Automatic differentiation in machine learning: a survey.
\newblock {\em Journal of Marchine Learning Research}, 18:1--43, 2018.

\bibitem{Julia-2017-a}
J.~Bezanson, A.~Edelman, S.~Karpinski, and V.~B. Shah.
\newblock Julia: A fresh approach to numerical computing.
\newblock {\em SIAM {R}eview}, 59(1):65--98, 2017.

\bibitem{adifor}
C.~Bischof, P.~Khademi, A.~Mauer, and A.~Carle.
\newblock {ADIFOR} 2.0: Automatic differentiation of {F}ortran 77 programs.
\newblock {\em IEEE Computational Science and Engineering}, 3(3):18--32, 1996.

\bibitem{Miletus.jl-2019}
S.~Byrne.
\newblock Miletus: Writing financial contracts in julia, 2019.

\bibitem{chen2018neural}
T.~Q. Chen, Y.~Rubanova, J.~Bettencourt, and D.~K. Duvenaud.
\newblock Neural ordinary differential equations.
\newblock In {\em Advances in Neural Information Processing Systems}, pages
  6571--6583, 2018.

\bibitem{Gen.jl-2019}
M.~F. Cusumano-Towner, F.~A. Saad, A.~K. Lew, and V.~K. Mansinghka.
\newblock Gen: A general-purpose probabilistic programming system with
  programmable inference.
\newblock In {\em Proceedings of the 40th ACM SIGPLAN Conference on Programming
  Language Design and Implementation}, PLDI 2019, pages 221--236, New York, NY,
  USA, 2019. ACM.

\bibitem{de2018end}
F.~de~Avila Belbute-Peres, K.~Smith, K.~Allen, J.~Tenenbaum, and J.~Z. Kolter.
\newblock End-to-end differentiable physics for learning and control.
\newblock In {\em Advances in Neural Information Processing Systems}, pages
  7178--7189, 2018.

\bibitem{Degrave_2019}
J.~Degrave, M.~Hermans, J.~Dambre, and F.~wyffels.
\newblock A differentiable physics engine for deep learning in robotics.
\newblock {\em Frontiers in Neurorobotics}, 13, Mar 2019.

\bibitem{blas}
J.~J. Dongarra, J.~D. Cruz, S.~Hammarling, and I.~S. Duff.
\newblock Algorithm 679: A set of level 3 basic linear algebra subprograms:
  model implementation and test programs.
\newblock {\em ACM Transactions on Mathematical Software (TOMS)}, 16(1):18--28,
  1990.

\bibitem{2019arXiv190401681D}
E.~{Dupont}, A.~{Doucet}, and Y.~{Whye Teh}.
\newblock {Augmented Neural ODEs}.
\newblock {\em arXiv e-prints}, page arXiv:1904.01681, Apr 2019.

\bibitem{XLA.jl-2018}
K.~Fischer and E.~Saba.
\newblock Automatic full compilation of {J}ulia programs and {ML} models to
  cloud {TPU}s.
\newblock {\em CoRR}, abs/1810.09868, 2018.

\bibitem{xla.jl}
K.~Fischer and E.~Saba.
\newblock {XLA}.jl: Compiling {J}ulia to {XLA}.
\newblock \url{https://github.com/JuliaTPU/XLA.jl}, 2018.

\bibitem{2019sbc}
D.~Gandhi, M.~Innes, E.~Saba, K.~Fischer, and V.~Shah.
\newblock {J}ulia {E} {F}lux: {M}odernizando o {A}prendizado de {M}áquina,
  2019.

\bibitem{10.1093/bioinformatics/btn278}
P.~Gao, A.~Honkela, M.~Rattray, and N.~D. Lawrence.
\newblock {Gaussian process modelling of latent chemical species: applications
  to inferring transcription factor activities}.
\newblock {\em Bioinformatics}, 24(16):i70--i75, 08 2008.

\bibitem{Turing.jl-2018}
H.~Ge, K.~Xu, and Z.~Ghahramani.
\newblock Turing: Composable inference for probabilistic programming.
\newblock In {\em International Conference on Artificial Intelligence and
  Statistics, {AISTATS} 2018, 9-11 April 2018, Playa Blanca, Lanzarote, Canary
  Islands, Spain}, pages 1682--1690, 2018.

\bibitem{Measurements.jl-2016}
M.~{Giordano}.
\newblock {Uncertainty propagation with functionally correlated quantities}.
\newblock {\em ArXiv e-prints}, Oct. 2016.

\bibitem{doi:10.1137/S0363012902419059}
E.~Gobet and R.~Munos.
\newblock Sensitivity analysis using {I}tô--{M}alliavin calculus and
  martingales, and application to stochastic optimal control.
\newblock {\em SIAM Journal on Control and Optimization}, 43(5):1676--1713,
  2005.

\bibitem{DBLP:journals/corr/abs-1810-01367}
W.~Grathwohl, R.~T.~Q. Chen, J.~Bettencourt, I.~Sutskever, and D.~K. Duvenaud.
\newblock {FFJORD:} free-form continuous dynamics for scalable reversible
  generative models.
\newblock {\em CoRR}, abs/1810.01367, 2018.

\bibitem{8062736}
D.~{Hartman} and L.~K. {Mestha}.
\newblock A deep learning framework for model reduction of dynamical systems.
\newblock In {\em 2017 IEEE Conference on Control Technology and Applications
  (CCTA)}, pages 1917--1922, Aug 2017.

\bibitem{He2016DeepRL}
K.~He, X.~Zhang, S.~Ren, and J.~Sun.
\newblock Deep residual learning for image recognition.
\newblock {\em 2016 IEEE Conference on Computer Vision and Pattern Recognition
  (CVPR)}, pages 770--778, 2016.

\bibitem{Hochreiter:1997:lstm}
S.~Hochreiter and J.~Schmidhuber.
\newblock Long short-term memory.
\newblock {\em Neural Comput.}, 9(8):1735--1780, Nov. 1997.

\bibitem{latent}
Y.~Hu, S.~Boker, M.~Neale, and K.~Klump.
\newblock Coupled latent differential equation with moderators: Simulation and
  application.
\newblock {\em Psychological methods}, 19, 05 2013.

\bibitem{huang2000malliavin}
Z.-y. Huang and J.-a. Yan.
\newblock Malliavin calculus.
\newblock In {\em Introduction to Infinite Dimensional Stochastic Analysis},
  pages 59--112. Springer, 2000.

\bibitem{Zygote.jl-2018}
M.~Innes.
\newblock Don't unroll adjoint: Differentiating {SSA}-form programs.
\newblock {\em CoRR}, abs/1810.07951, 2018.

\bibitem{rlvsdp}
M.~Innes, N.~M. Joy, and T.~Karmali.
\newblock Reinforcement learning vs. differentiable programming.
\newblock \url{https://fluxml.ai/2019/03/05/dp-vs-rl.html}, 2019.

\bibitem{Flux.jl-2018}
M.~Innes, E.~Saba, K.~Fischer, D.~Gandhi, M.~C. Rudilosso, N.~M. Joy,
  T.~Karmali, A.~Pal, and V.~Shah.
\newblock Fashionable modelling with {F}lux.
\newblock {\em CoRR}, abs/1811.01457, 2018.

\bibitem{jax}
M.~Johnson, R.~Frostig, D.~Maclaurin, and C.~Leary.
\newblock {JAX}: Autograd and xla.
\newblock \url{https://github.com/google/jax}, 2018.

\bibitem{jones2003write}
S.~P. Jones and J.-M. Eber.
\newblock How to write a financial contract.
\newblock 2003.

\bibitem{jones2000composing}
S.~P. Jones, J.-M. Eber, and J.~Seward.
\newblock Composing contracts: an adventure in financial engineering.
\newblock {\em ACM SIG-PLAN Notices}, 35(9):280--292, 2000.

\bibitem{gordon-bell-2018-a}
T.~Kurth, S.~Treichler, J.~Romero, M.~Mudigonda, N.~Luehr, E.~Phillips,
  A.~Mahesh, M.~Matheson, J.~Deslippe, M.~Fatica, et~al.
\newblock Exascale deep learning for climate analytics.
\newblock In {\em Proceedings of the International Conference for High
  Performance Computing, Networking, Storage, and Analysis}, page~51. IEEE
  Press, 2018.

\bibitem{li2018differentiable}
T.-M. Li, M.~Aittala, F.~Durand, and J.~Lehtinen.
\newblock Differentiable monte carlo ray tracing through edge sampling.
\newblock In {\em SIGGRAPH Asia 2018 Technical Papers}, page 222. ACM, 2018.

\bibitem{doi:10.1243/09544062JMES683}
K.~Ordaz-Hernandez, X.~Fischer, and F.~Bennis.
\newblock Model reduction technique for mechanical behaviour modelling:
  Efficiency criteria and validity domain assessment.
\newblock {\em Proceedings of the Institution of Mechanical Engineers, Part C:
  Journal of Mechanical Engineering Science}, 222(3):493--505, 2008.

\bibitem{pearlmutter2008reverse}
B.~A. Pearlmutter and J.~M. Siskind.
\newblock Reverse-mode {AD} in a functional framework: Lambda the ultimate
  backpropagator.
\newblock {\em ACM Transactions on Programming Languages and Systems (TOPLAS)},
  30(2):7, 2008.

\bibitem{PyTorchYear}
{PyTorch Team}.
\newblock {PyTorch}, a, year in...
\newblock \url{pytorch.org/blog/a-year-in}, 2018.
\newblock Accessed: 2018-09-22.

\bibitem{pytorch}
{PyTorch Team}.
\newblock The road to 1.0: production ready {PyTorch}.
\newblock \url{https://pytorch.org/blog/a-year-in/}, 2018.
\newblock Accessed: 2018-09-22.

\bibitem{DBLP:journals/corr/abs-1902-02376}
C.~Rackauckas, M.~Innes, Y.~Ma, J.~Bettencourt, L.~White, and V.~Dixit.
\newblock Diffeqflux.jl - {A} julia library for neural differential equations.
\newblock {\em CoRR}, abs/1902.02376, 2019.

\bibitem{2018arXiv181201892R}
C.~{Rackauckas}, Y.~{Ma}, V.~{Dixit}, X.~{Guo}, M.~{Innes}, J.~{Revels},
  J.~{Nyberg}, and V.~{Ivaturi}.
\newblock {A Comparison of Automatic Differentiation and Continuous Sensitivity
  Analysis for Derivatives of Differential Equation Solutions}.
\newblock {\em arXiv e-prints}, page arXiv:1812.01892, Dec 2018.

\bibitem{DifferentialEquations.jl-2017}
C.~Rackauckas and Q.~Nie.
\newblock Differentialequations.jl – a performant and feature-rich ecosystem
  for solving differential equations in julia.
\newblock 5(1), 2017.
\newblock Exported from https://app.dimensions.ai on 2019/05/05.

\bibitem{Rackauckas2017ADAPTIVEMF}
C.~V. Rackauckas and Q.~Nie.
\newblock Adaptive methods for stochastic differential equations via natural
  embeddings and rejection sampling with memory.
\newblock {\em Discrete and continuous dynamical systems. Series B}, 22
  7:2731--2761, 2017.

\bibitem{ROMEROUGALDE2013170}
H.~M.~R. Ugalde, J.-C. Carmona, V.~M. Alvarado, and J.~Reyes-Reyes.
\newblock Neural network design and model reduction approach for black box
  nonlinear system identification with reduced number of parameters.
\newblock {\em Neurocomputing}, 101:170 -- 180, 2013.

\bibitem{wang2018demystifying}
F.~Wang, X.~Wu, G.~Essertel, J.~Decker, and T.~Rompf.
\newblock Demystifying differentiable programming: Shift/reset the penultimate
  backpropagator.
\newblock {\em arXiv preprint arXiv:1803.10228}, 2018.

\bibitem{zhang_2004}
H.~Zhang.
\newblock {\em The Malliavan Calculus}.
\newblock PhD thesis, 2004.

\bibitem{yao.jl}
X.~zhe Luo, J.~guo Liu, P.~Zhang, and L.~Wang.
\newblock Yao.jl: Extensible, efficient quantum algorithm design for humans.
\newblock In preparation, 2019.

\bibitem{pmlr-v5-alvarez09a}
M.~Álvarez, D.~Luengo, and N.~D. Lawrence.
\newblock Latent force models.
\newblock In D.~van Dyk and M.~Welling, editors, {\em Proceedings of the Twelth
  International Conference on Artificial Intelligence and Statistics}, volume~5
  of {\em Proceedings of Machine Learning Research}, pages 9--16, Hilton
  Clearwater Beach Resort, Clearwater Beach, Florida USA, 16--18 Apr 2009.
  PMLR.

\end{thebibliography}
\bibliographystyle{abbrv}

\end{document}